# Frequency-domain "single-shot" (FDSS) transient absorption spectroscopy using a variable-length grating pair compressor. [a)]

Ilya A. Shkrob [*], Dmitri A. Oulianov, and Robert A. Crowell

*Chemistry Division, Argonne National Laboratory, Argonne, IL 60439*

**Abstract**

Single-shot ultrafast spectroscopy based on the frequency encoding of transient absorbance kinetics (FDSS) is demonstrated. These kinetics are sampled spectrally using linearly chirped pulses derived from a Ti:sapphire laser. A variable length grating pair compressor is used to provide group velocity dispersion out to -1.6 $ps^2$ and achieve the sampling of 512 channels per a 2-to-160 ps window with sensitivity > $5 \times 10^{-4}$. The possibilities of FDSS are illustrated with studies of three photon ionization of liquid water and one-photon excitation of a thin-film amorphous Si:H semiconductor.

___________________________________________________________

[*] To whom correspondence should be addressed: *Tel* 630-2529516, *FAX* 630-2524993, *e-mail:* shkrob@anl.gov.

I. INTRODUCTION

This work completes a series of papers on frequency domain "single-shot" transient absorption spectroscopy (FDSS) originally suggested by Beddard et al. [1] and further developed at our laboratory. [2,3] Our aim is to demonstrate that this technique is as general and versatile as the familiar pump-probe spectroscopy (PPS), but surpasses the latter in the averaging efficiency. We emphasize that the "single-shot" aspect of the FDSS spectroscopy relates to the fact that the entire transient absorbance (TA) kinetics are sampled for each probe pulse; in fact, many of such kinetics may need be averaged to obtain the desirable signal-to-noise (S/N) ratio. The single shot nature of this spectroscopy distinguishes it from the PPS in which the kinetics are collected in a point-to-point fashion, by stepping the delay time of the probe pulse relative to that of the pump pulse. By its very nature, FDSS is less sensitive to shot-to-shot laser and sample instabilities, and this feature is the chief (but not the only) advantage of this technique. Another important advantage is the elimination of the time-consuming mechanical movement of the delay line, which also improves averaging efficiency.

Like several other spectroscopy techniques based on frequency encoding of temporal information (see refs. 2 and 3 for a brief review), FDSS imprints the kinetics onto the phase and, eventually, the spectrum of the probe light. A linear chirp is introduced into a Gaussian probe pulse centered at the frequency $\omega_0$ so that the frequency-dependent phase $\phi(\omega)$ of this pulse is given by $\phi(\omega) \approx 1/2 \; \phi''(\omega_0) \; \Delta\omega^2$, where $\Delta\omega = \omega - \omega_0$ is the frequency offset and $\phi''(\omega_0)$ is the group velocity dispersion (GVD). This probe pulse is delayed in time by $T$ relative to the pump pulse, passed through a sample, and the spectrum of the transmitted light is analyzed and normalized by the spectrum of a reference beam derived from the same probe light before it interacts with the sample. The FDSS signal $S(\omega) = -\Delta T_\omega / T_\omega$, where $T_\omega$ is the transmission coefficient of the quasimonochromatic component of the probe light with frequency $\omega$, and $\Delta T_\omega$ is the photoinduced change in this transmission coefficient. Since some of these components interact with the sample later than others, the TA kinetics is imprinted on the spectrum of the transmitted light. As shown in ref. 2, the FDSS signal $S(\omega)$ follows the TA kinetics $\Gamma(t)$, $S(\omega) \approx \Gamma(t = T_e(\omega))$, where the group delay time $T_e$ is given by

$$T_e(\omega) = T - \phi'(\omega) \approx T - \phi''(\omega_0)\Delta\omega \qquad (1)$$

Due to the interference of different quasimonchromatic components in the probe beam, the correspondence between these FDSS kinetics $S(\omega)$ and the TA kinetics $\Gamma(t)$ is violated near the kinetic origin ($T_e = 0$), where $S(\omega)$ strongly oscillates.[2] The period of these oscillations is of the same order of magnitude as $\tau_{GVD} = \sqrt{|\phi''(\omega_0)|}$. The oscillation pattern strongly depends on the phase of the complex refraction index of photoinduced species,[1-3] and this phase may be obtained by the analysis of this pattern.[2,3] Thus, FDSS combines the features of the TA spectroscopy and chirped pulse interferometry.[3] Since the GVD can be changed over several orders of magnitude by changing the distance between the gratings in the grating pair compressor,[4] the time window accessible to FDSS spectroscopy is widely tunable, from a few picoseconds to hundreds of picoseconds.[2,3,4] This sampling window is given by $|s|\tau_p$, where $s = \phi''(\omega_0)/\tau_p^2$ is the stretch/compression factor ($s < 0$ for compression) and $\tau_p$ is the $1/e$ width of the Gaussian probe pulse (before chirping). With the typical compressor designs that are used for chirped-pulse amplification (CPA),[4] "single-shot" kinetic measurements over 300-500 ps are possible. In Sec. II of this paper and in ref. 3, we describe a tandem PPS-FDSS setup based on a variable-length grating compressor, in which the time window of the observation can be continuously changed between 2 and 160 ps.

In the present study, we demonstrate the advantages of FDSS for two photosystems: liquid water flowed in a 160 μm thick high-speed jet ionized by absorption of three or more 400 nm photons (Sec. III.A) and a 1.3 μm thin film of amorphous Si:H semiconductor in which charge separation is induced by 400 nm light. Two other examples of the use of FDSS are given in ref. 3. It is shown in the present work that (i) the FDSS technique yields superior signal-to-noise (S/N) ratio to that of the PPS for the same acquisition time in a situation when amplitude variations are 50-100% of the TA signal and (ii) this spectroscopy is suitable both for thick and thin samples, including films that exhibit well-resolved interference fringes at the probe wavelengths.

## II. EXPERIMENTAL.

A 1.3 μm thick film of undoped amorphous (*a*-) Si:H alloy (8 at. % of H) deposited on 1 mm suprasil substrate was obtained from Prof. H. Fritzsche of the University of Chicago (see ref. 5 for more detail). Pure water with conductivity < $2\times10^{-9}$ $\Omega^{-1}$ cm$^{-1}$ was flowed using a 160 μm thick high-speed jet. The linear velocity of the fluid was > 1 m/s. All measurements were carried out at 295 K.

The diagram of the setup is shown in Fig. 1; more details are given in ref. 3. The pump and probe pulses were derived from a femtosecond Ti:sapphire system. 800 nm pulses from the oscillator were stretched to 80 ps in a 1200 *g*/mm grating stretcher. These 2 nJ pulses were then amplified to 4 mJ in a two-stage multipass Ti:sapphire amplifier. The repetition rate of the amplifier was 1 kHz and the pulse to pulse stability was typically 3%. The output from the amplifier was split 1:20. The main part of the beam was passed through a grating compressor that yielded Gaussian probe pulses of 50 fs full width at half maximum (FWHM) and 3 mJ centered at 800 nm. These pulses were delayed using a retroreflector RR1 on a double path 1 m motorized translation stage TS2 and doubled in frequency using a 400 μm *β*-BaB$_2$O$_4$ crystal (SHG in Fig. 1). The typical maximum output of the 400 nm light was 80 μJ (100-200 fs FWHM).

The probe pulse was chirped using a variable length grating pair compressor (1200 *g*/mm) in which the larger (11 cm wide) grating was mounted on a motorized translation stage TS1. This design gave us up to 55 cm of the slanted distance between the gratings for compression (that corresponds to maximum GVD of -1.6 ps$^2$). The spectrum of the compressed beam was continuously monitored using a miniature fiber optics spectrometer (FBO).The GVD was obtained from eq. (1).

The main part of the compressed beam was passed through a 50% beam splitter (BS). One beam was used as a reference, another as a probe. Before this beam splitter, the 800 nm beam was passed through a 2-3 mm aperture (A) and attenuated down to < 1 nJ. To eliminate the possibility of coherent artifacts, the pump and probe pulses were perpendicularly polarized. The 800 and 400 nm beams were focused with achromatic lenses L1 and L2 and overlapped in the sample (SM) at 6.5º. The transmitted probe was collimated using a thin lens L3, and the "signal" and "reference" beams were used for either the standard pump-probe or for FDSS detection (see ref. 3 for more detail). The vertical bars shown in the PPS kinetics given in Figs. 2(b), 3(a), and 4 represent 95% confidence limits for each data point.

In the FDSS experiment, the probe and reference beams were focussed on a diffuser (DF) and the monochromator slit with a cylindrical lens (CL). With an $f = 27$ cm monochromator (SPEX model 270M) equipped with a 1200 $g$/mm grating, a spectral resolution of 2-5 cm$^{-1}$ was achieved for a 50-200 μm slit opening. A cooled 512 channel dual diode array (Princeton Instruments DIDA512) was used for detection. The typical duty cycle of this detector was 70%, with 60 ms exposure to the laser light per frame. Several such frames were averaged, and series with the pump pulse off and on were alternated. A mechanical shutter (SH2 in Fig. 1) was used to block the pump light. The ratios of the probe and the reference signals were calculated for each acquisition channel, and the "pump on" ratios were normalized by the "pump off" ones to obtain the optical density $\Delta OD_\omega = -\log(T_{on}/T_{off}) \approx 2.3\, S(\omega)$ for each channel. 100-500 such "on/off" series were averaged to obtain the kinetic profiles shown in the next section. Before the experiment, "pump on" and "pump off" dark signals with the probe light blocked by another shutter (SH1 in Fig. 1) were collected and subtracted from the spectra. In a typical "blank" experiment with the pump beam blocked, the "on" and "off" series for the total of 3x10$^4$ pulses converged to an optical density of 2x10$^{-5}$ and the standard deviation across the spectrum was 5x10$^{-4}$.

## III. RESULTS AND DISCUSSION.

### A. Photoionization of neat water.

The conduction band of liquid water is positioned 8-9 eV above the ground state and simultaneous absorption of three 400 nm (3.1 eV) photons is sufficient to ionize this liquid.[6] In the course of this ionization, an electron is injected into the solvent bulk ca. 1 nm away from the parent molecule; these "hot" electrons rapidly (< 5 ps) thermalize yielding the so-called "hydrated electron", $e_{aq}^-$, that has a broad absorption band across the 300-1300 nm range, centered at 720 nm (at 25ºC).[7] This band corresponds to the bound-to-bound ($1s \rightarrow 2p$) transition of the electron localized in a spherical solvent cavity.[7] Since the absorption band of this hydrated electron is much broader than the spectral width of the 800 nm probe pulse, this spectrum is flat over the sampling range and shows little time evolution after the first few picoseconds.

A three 400 nm photon absorption coefficient for water of 900 cm$^3$/TW$^2$ has been reported;[8] recently, it has been revised to 270 cm$^3$/TW$^2$ and a quantum yield of 0.41 hydrated electrons per 3 photons absorbed was estimated.[9] When the radiance of 400 nm photons exceeds 0.6-1.2 TW/cm$^2$, another 400 nm photon may be absorbed by the electron which is then promptly injected deep into the conduction band of the solvent (Fig. 2(a)). In less than 200 fs, this electron localizes several nanometers away from the parent hole.[10] As a result, the "3+1" photoexcitation process makes the geminate recombination less efficient.[6,9,10] The fraction of hydrated electrons that escape geminate recombination with the hole at $t$=500 ns increases from 70% to 92% (Fig. 2(b)) as the average width of the electron-hole distribution increases from 1.15 nm (for the 3-photon process) to 2.7 nm (for the "3+1" process).[6]

We chose this photosystem because it provides one of the worst scenarios for PPS: The high, mixed photon order of the ionization causes strong shot-to-shot variation in the electron yield. This yield varies strongly across the excitation beam and along the path of the 400 nm light. For an average pump radiance of 0.1-0.4 TW/cm$^2$, a switchover of the dynamic behavior from 3 photon to the "3+1" process occurs at the higher end of this range. Since water breaks down at 1-5 TW/cm$^2$, the photolysis is carried out in a high-speed jet, and the light scatter and thermal lensing in the jet further add to the noise. The absorption of the 400 nm light in water is extremely non-uniform; e.g., at 0.5-1 TW/cm$^2$, 80-90% of the light is absorbed in a 10-20 μm thin layer near the surface. As a consequence, the temperature of water in this layer instantly increases to 30-80 ºC,[9] and this sudden temperature jump further destabilizes the TA signal since the absorption spectrum of hydrated electron shifts to the red in hot water.[7] Although $\Delta OD$ of 0.01-1 can be obtained for radiances just below the dielectric breakdown (at 10 TW/cm$^2$), the shot-to-shot variance of the TA is comparable to the TA signal itself.

Pump-probe kinetics of the hydrated electron in neat H$_2$O were obtained in the 3-photon regime using 0.15 TW/cm$^2$, 400 nm pump pulses of 200 fs FWHM (Fig. 3(a), open circles). The solid line in Fig. 3(a) indicates FDSS kinetics from the same photosystem under identical excitation conditions. For the compression factor $s = -2160$, the maximum time window is ca. 60 ps ($\tau_{GVD}$ of 0.9 ps) and the decay kinetics of $e^-_{aq}$ were quite flat (see Fig. 2(b)). The oscillation pattern near the kinetic origin is similar to the one found for hydrated electron in the iodide photosystem;[3] perhaps, its characteristic asymmetric profile also originates in the solvation dynamics of pre-thermalized electron on the short time scale.[3] Fig. 3(b) shows FDSS kinetics obtained for a greater GVD of -1.46 ps$^2$ for three pump powers in the "3+1" region; the window is extended to 140 ps. As $\Delta OD$ increases from 0.017 to 0.2 to 0.4, the amplitude of the oscillations decreases.

Similar trends were observed for other photosystems studied (ref. 3): the interference of quasimonochromatic components that give rise to the oscillation pattern in Fig. 3 is weaker in the strongly absorbing samples.

Note that same "noise" level was observed in the FDSS kinetics before and after the pump pulse (Fig. 3), suggesting that most of the "noise" was due to the unstable flow of the sample rather than pump variation. We believe that most of the scatter in the FDSS traces is caused by an unstable jet surface: Near the nozzle, the jet is a concave lens that strongly refracts and disperses the probe light, and sinous and capillary waves and other flow disturbances constantly modify the spectrum of the transmitted light. Though the S/N ratio for the FDSS kinetics obtained using the high-speed jet was worse than the same ratio obtained in ref. 3 for solid samples and liquid samples flowed in a glass cell, the averaging time needed to achieve this S/N ratio was 10 times shorter for FDSS than for the PPS experiment, under identical excitation conditions.

## B. Thin-film amorphous Si:H alloy.

When the sample thickness $d$ is comparable to the wavelength of the probe light, the sample will exhibit well-resolved interference fringes. As a result, the PPS measurement of the time-dependent photoinduced change $\Delta\varepsilon_\omega(t) = \Delta\varepsilon_\omega \Gamma(t)$ in the dielectric function $\varepsilon_\omega$ becomes quite complicated.[5,11,12] Such a situation frequently occurs in studies of thin film semiconductors.[12] For a quasimonochromatic component of the probe light with frequency $\omega$, the change $\Delta T_\omega$ in the transmission coefficient $T_\omega$ induced by the pump pulse is given by

$$\Delta T_\omega / T_\omega = 2 \ \mathrm{Re} \ \tau_\omega \ \Delta\varepsilon_\omega(t), \tag{2}$$

where we introduced the complex spectral response coefficient $\tau_\omega = t_\omega^{-1}(dt/d\varepsilon)_\omega$ and $t_\omega$ is the (complex) Fresnel transmission coefficient for light of the frequency $\omega$. For a very thin sample ($k_\omega d \ll 1$, where $k_\omega = \omega\sqrt{\varepsilon_\omega}/c$ is the wave vector), the interference fringes are well spaced and the PPS signal $-\Delta T_\omega/T_\omega \propto \Delta\varepsilon''_\omega(t)$. For a thicker sample, the complex factor $\tau_\omega$ rapidly oscillates with frequency $\omega$ as $\exp(\pm ik_\omega d)$, and in order to extract $\Delta\varepsilon_\omega(t)$ from the PPS kinetic traces one needs to (i) determine this factor in a separate experiment, (ii) measure transient reflection of the sample in addition to the transmission, and (iii) invert linear equations that express these two signals vs. $\Delta\varepsilon'_\omega(t)$ and $\Delta\varepsilon''_\omega(t)$.[12,13] Involved as it is, this procedure (known as the "inversion method")[12] is insufficient to obtain the real part of the dielectric function near the fringe extrema, where $\mathrm{Im}\,\tau_\omega = 0$. On the other hand, precisely due to the latter condition, $\Delta\varepsilon''_\omega(t)$ can be determined from a single TA measurement near these fringe extrema, even if $\Delta\varepsilon'_\omega(t)$ is large, provided that the fringe period in the frequency domain is much greater than the spectral width of the probe pulse, $1/\tau_p$.[5,13] The retrieval of $\Delta\varepsilon_\omega(t)$ is further complicated by inhomogeneous absorption of the pump that changes the inversion matrix. Moon and Tauc[12] found that this method completely breaks down for large TA signals ($|\Delta T/T| > 0.01$). Even when the TA signal is weak, the results are very sensitive to small errors in $\tau_\omega$,[12] which is seldom known with the required accuracy for the exact spot probed with the laser light (e.g., compare traces (iii) in Figs. 5 and 6 given below for two spots on the same sample). As shown below, FDSS is, in some ways, preferable to the inversion method. The advantage of FDSS is that it inherently combines spectral and kinetic measurement in a single experiment whereas the inversion relies on several independent measurements.

In ref. 2 it was shown that for a flat sample of thickness $d$ that uniformly absorbs the pump light, the signal $S(\omega)$ is given by

$$S(\omega) \approx -2 \, \text{Re}\left\{\int_{-\infty}^{+\infty} d\Omega \; K_{\Omega-\omega} \; e^{i(\Omega-\omega)T} \; E_\Omega/E_\omega \; \Theta(\Omega)\right\}, \qquad (3)$$

where $K_\mu$ is the Fourier component of TA kinetics $\Gamma(t)$, $E_\omega \propto \exp\left[-i\Delta\omega^2 \tau_p^2/2 - i\phi(\omega)\right]$ is the Fourier transform of the electric field of the probe light, and the function $\Theta(\Omega)$ is defined as

$$\Theta(\Omega) \approx \Delta k_\omega \; t_\omega^{-1} \; (t_\Omega - t_\omega)/(k_\Omega - k_\omega), \qquad (4)$$

where $\Delta k_\omega$ is the photoinduced change in the wave vector. For $\phi(\omega) = 0$ (no chirp), eq. (3) gives the PPS signal as a function of the delay time $T$ of the pump for the quasimonochromatic component with frequency $\omega$. If the probe light is not analyzed before the detection, as is the case in most PPS experiments, this expression should be averaged over the band width of the probe light. If this band is much narrower than the fringe period (i.e., $\omega\tau_p \gg k_\omega d$), eq. (2) is obtained.

If the sample is very thin ($k_\omega d \ll 1$), then $\Theta(\Omega) \approx \Theta(\omega) \approx \tau_\omega \Delta\varepsilon_\omega$ is a slow function of frequency $\omega$. In this thin-film limit, the phase of the FDSS signal and the oscillation pattern near the kinetic origin (see, for example, Fig. 3) is given by the phase of the product $\tau_\omega \Delta\varepsilon_\omega$, similarly to the case of PPS kinetics given by eq. (2). Since the position of the kinetic origin (where $T_e(\omega) = 0$, eq. (1)) in the frequency domain $\omega$ depends on the delay time $T$ of the pump, different patterns are obtained as this origin is placed, by changing the delay time, at different positions with respect to the fringe pattern. This occurs due to the frequency dependence of the phase of the complex function $\tau_\omega$. For thicker samples, the fringe period may easily be comparable to the band width of the probe pulse ($\approx 2/\tau_p$), and the exact solution, eq. (4), should be used. For any sample, the FDSS signal $S(\omega)$ asymptotically converges to $T(\omega) = -2 \, \text{Re} \, \tau_\omega \exp(i\phi_\varepsilon)$ at long group delay times $T_e$, $S(\omega)/\mathcal{T}(\omega) \approx |\Delta\varepsilon_\omega|\ \Gamma(T_e)$, where $\phi_\varepsilon$ is the phase of $\Delta\varepsilon_\omega$. Thus, if the TA kinetics are flat (i.e., $\Gamma(|s|\tau_p) \approx 1$), function $\mathcal{T}(\omega)$ can be obtained by delaying the probe pulse in time so that the kinetic origin (where $T_e(\omega) = 0$) is outside of the spectrometer window, $\omega_0 \pm 1/\tau_p$ (see below).

To demonstrate the peculiarities of FDSS for thin samples, we have chosen amorphous hydrogenated silicon (*a*-Si:H) which is a commercial thin-film material for solar energy conversion. [5,11-19] This material has an optical gap of 1.75 eV and is transparent at our probe wavelength, 800 nm. The dynamics of photoinduced free carriers and trapped charges in *a*-Si:H has been extensively studied (see reviews in refs. [5, 11, and 14]). Upon short-pulse excitation with < 600 nm photons, free carriers are injected in their respective bands. These carriers thermalize, [15,16,17] recombine with each other (with rate constant of 2.3x10$^{-8}$ cm$^3$/s) [5,15,16,17] and descend into shallow (60-100 meV) traps (~10$^{20}$ cm$^{-3}$) with a rate constant of (1-2)x10$^{-8}$ cm$^3$/s. [5,18,19] The scattering time of the plasma is very short, ca. 0.5 fs, [11,14] and no TA signal from the free carriers was observed in the visible and near infra-red for carrier density < 10$^{19}$-10$^{20}$ cm$^{-3}$. [5,13,14,18,19] For initial carrier densities of 10$^{17}$-10$^{18}$ cm$^{-3}$, the TA signal is dominated by band-tail charges that slowly recombine (with a rate constant of 6x10$^{-9}$ cm$^3$/s), [5] by hopping and thermal emission, and descend into < 10$^{17}$ cm$^{-3}$ of deep traps (such as dangling Si bonds). [5,11,12,18,19] The intraband absorption of these trapped charges is a smooth, featureless curve that gradually ascends from the visible into the near infra-red. [18,19] At low carrier density, the decay kinetics of $\Delta\varepsilon'_\omega(t)$ and $\Delta\varepsilon''_\omega(t)$ are slow (lasting hundreds of ps) and dispersive. [5,11,12] $\Delta\varepsilon'_\omega(t)$ is negative, and the initial phase (for *t* <100 ps) is close to 107° (at 1033 nm). [18,19]

Fig. 4 shows PPS kinetics obtained for 400 nm excitation of a $d$=1.3 μm thick film of $a$-Si:H. At this excitation energy, the pump light is absorbed in 30 nm layer near the surface, resulting in high initial density (>$10^{19}$ cm$^{-3}$) and considerable excess energy (ca. 1.35 eV) of the photocarriers. The center frequency $\omega_0$ of the probe pulse, 800 nm, is matched with the transmission maximum of the film at 796 nm, and the TA signal is dominated by photoinduced absorbance from the free carriers and trapped charges.[5] The short-lived "spike" (< 5 ps) near the kinetic origin has nearly the same decay profile as the (positive) $\Delta\varepsilon'_\omega(t)$ and (negative) $\Delta\varepsilon''_\omega(t)$ in the 310 nm pump - 310 nm probe experiment by Wraback et al.[16] Similar PPS kinetics with a life time of 1.5 ps for the spike were observed in a 400 nm pump - 2.86 μm probe experiment at Argonne (unpublished). Tauc and coworkers [11,16,17] give an estimate of 2 eV/ps for the rate of carrier relaxation in a-Si:H, which gives 1.5 ps for thermalization time after 400 nm photoexcitation. The decay rate of the "spike" changes with the pump intensity, and the kinetics can be interpreted in terms of a monoexponential process with time constant of ca. 2 ps (that Wraback et al.[16,17] associate with carrier relaxation) and a bimolecular process with rate constant of 4x10$^{-10}$ cm$^3$/s (presumably, due to recombination of these "hot" carriers). For $t$ >5 ps, the PPS kinetics show slow decay over > 1 ns; with only a few per cent drop in the photoinduced optical density $\Delta OD$ over the first 50 ps after the photoexcitation pulse. From these results, it may be expected that picosecond FDSS kinetics would be flat. Such dispersive TA kinetics are very typical for amorphous semiconductors: one can always find a time window where these kinetics are nearly flat. Thus, for a suitably long pump delay, the wavelength dependence of the coefficient $\mathcal{T}(\omega)$ can be determined as discussed above. For the thin-film system examined, this function has been obtained at $T$=300 ps (Figs. 5 and 6, traces (iii)). Following other authors, we assume that the photoinduced change in the dielectric function is constant over the narrow spectral band of the probe pulse,[18,19] and the wavelength dependence of $\mathcal{T}(\omega)$ is due to the wavelength-dependent spectral function $\tau_\omega$ alone.[12]

In Figs. 5 and 6(a), FDSS kinetics for the same sample are given at several delay times $T$ of the pump pulse and two compression factors, $s$=-630 ($\tau_{GVD} = 0.53\,fs$) and $s$=-3780 ($\tau_{GVD} = 1.23\,fs$). In Fig. 5(b), $T$=0 ps and $T$=9 ps kinetics $S(\omega)$ obtained for $s$=-630 (shown in Fig. 5(a)) were normalized by the $T$=300 ps kinetics that yield the spectral response function $\mathcal{T}(\omega)$. These normalized kinetics are flat after the first few picoseconds, suggesting that the normalization procedure succeeds in compensating for the curved transmission profile. In trace (i) of Fig. 6(b), the oscillation patterns obtained for $s = -630$ at two delay times $T$ of the pump are juxtaposed in the group delay time (given in the units of $\tau_{GVD}$). In the frequency domain, $T$=0 ps corresponds to the time origin placed at the transmittance maximum whereas $T$=9 ps corresponds to this origin placed at the reflectance maximum (Fig. 5(b)). Despite a considerable change in the phase of the complex factor $\tau_\omega$ for these two positions (see below), the two oscillation patterns are almost exactly the same. The same applies to the kinetic traces obtained for a greater compression factor (Fig. 6). This result is counterintuitive because, as explained above, in general the oscillation pattern exhibited by FDSS kinetics depends strongly on the interference of quasimonochromatic components, and this interference may be expected to change as the kinetic origin is scanned across the fringe pattern. This is certainly the case for very thin samples, for which the band width of the probe pulse is much smaller than the fringe period.

Below we demonstrate that the general theory of FDSS spectroscopy given in ref. 2 and eq. (3) derived therein account well for these observations. Fig. 7 shows the function $\mathcal{T}(\omega)$ calculated using Fresnel coefficients given by eqs. (A7) and (A8) in ref. 2 for $d=1.27$ μm and $\phi_\varepsilon=0^0$ and $90^0$ (the refraction indexes for the film and the glass substrate are taken from ref. 14). This function has a maximum exactly at the center frequency $\omega_0$ of the probe pulse, and the fringe period is close to the FWHM of this pulse. The calculation indicates that the phase of the complex factor $\tau_\omega$ in eq. (2) changes from 90º at the spectral center (where the transmission is minimum) to ±65º at the limits of the optimum spectral range (where the reflection is maximum/minimum).

Fig. 7 exhibits FDSS kinetics obtained by numerical solution of eq. (3) for $T=0$ (kinetic origin at the transmission minimum), $T=35$ ps (kinetic origin at the reflection maximum), and $T=20$ ps for a photoabsorption signal ($\phi_\varepsilon=90º$) that exponentially decays with life time of 40 ps. Although these simulated FDSS kinetics change considerably as a function of $T$, most of this change is in the weighting factor $\tau_\omega$. In Fig. 8(a), normalized signals $S(\omega)/\mathcal{T}'(\omega)$ are plotted; these normalized kinetics change very slightly for these three delay times. In Fig. 8(b), $S(\omega)/\mathcal{T}'(\omega)$ kinetics are plotted as a function of $T_e$ shifted by $T$. The latter shift is introduced in order to juxtapose the corresponding oscillation patterns. It is seen that the changes in the positions of crests are small, ca. 20-30% of what would be expected from eq. (3) in which one lets $\Theta(\Omega) \approx \tau_\omega \Delta\varepsilon_\omega$, as can be done in the thin-film limit (for $k_\omega d \ll 1$). E. g., for $T=20$ ps and 35 ps the spacing between the first pair of the oscillation crests is just 4.5% and 9.1% lower, respectively, than the same spacing for $T=0$ ps. This result is in full agreement with the experimental observations discussed above. Qualitatively, since the quasimonochromatic components of the chirped probe pulse cover the entire spectral range of the fringe, the phase shifts acquired by these components due to their interference in the sample cancel each other almost perfectly. In this sense, the resulting FDSS kinetics is much like the PPS kinetics, in which the change in the transmission of the probe light is averaged over the entire band of the probe pulse. The advantage of FDSS is that the spectral information needed to retrieve $\Delta\varepsilon_\omega$ is obtained simultaneously with these TA kinetics, in the same measurement.

We conclude that for thin-film samples that exhibit closely spaced interference fringes meaningful "TA kinetics" can be obtained from the FDSS traces by carrying out the $S(\omega)/\mathcal{T}(\omega)$ normalization. Like the corresponding PPS traces, these "TA kinetics" are a combination of transient absorption and reflection signals. Unlike PPS, FDSS simultaneously yields the spectral function $\mathcal{T}'(\omega)$ needed to obtain phase relations that are required for the kinetic analysis.

## IV. CONCLUSION.

Two photosystems were studied using FDSS in which the probe light was chirped using a variable length grating compressor. For these photosystems, FDSS yielded a S/N ratio superior to that of PPS over a shorter acquisition time. The sampling time was reduced due to the "single-shot" nature of the method and elimination of mechanical movement of the translation stage. The most dramatic reduction was for multiphoton excitation (Sec. III.A) since the amplitude variations were the strongest. TA kinetics for 3-to-4 photon ionization of water were obtained. The S/N ratio for these FDSS kinetics exceeded this ratio for PPS kinetics collected over 10 times longer acquisition time.

In Sec. III.B, we illustrated the use of FDSS for a thin-film sample that exhibited well resolved interference fringes whose spacing was similar to the spectral width of the probe pulse – an especially complex case for PPS [9] and "single-shot" spectroscopies based upon spatial encoding [20] or chirped-pulse interferometry. [20]

We conclude that the FDSS technique is widely applicable, versatile, and easy to implement experimentally. The frequency domain spectrometer can be built side-by-side with a pump-probe spectrometer in a single setup (Sec. II). The requirements for the compressor can be met with the standard equipment used for chirped-pulse amplification. The method does not require tight control over the profile of the probe and pump beam, as is the case with "single-shot" methods based on spatial encoding.[21]

## VII. ACKNOWLEDGMENT

We thank Drs. S. Pommeret, D. Gozstola, D. M. Bartels, and C. D. Jonah for many helpful discussions. This work was performed under the auspices of the Office of Basic Energy Sciences, Division of Chemical Science, US-DOE under contract No. W-31-109-ENG-38.

**Figure captions.**

**Fig. 1.**

The setup diagram. See Sec. II for more detail.

**Fig. 2.**

(a) A typical power dependence of $e_{aq}^-$ yield in 400 nm photoionization of neat liquid water in a 160 μm thick jet observed via the PPS-detected electron absorbance at 800 nm, ca. 16 ps after the 200 fs FWHM excitation pulse (at which time the electron thermalization is complete). The pump and probe beam radii were 56 and 14 μm, respectively. At low irradiance (< 0.5 TW/cm$^2$), the ionization is 3-photon and results in the geminate decay kinetics for which the escape yield of the electron is ca. 72%. At higher irradiance, the electron yield linearly increases with the pump power, indicating the occurrence of the "3+1" excitation process. Simultaneously, the time profile of the kinetics changes so that the escape yield approaches > 90%. (b) TA kinetics obtained in the "3+1" regime (for 80 μJ pump pulse). The decay is second order *(solid line)* and originates mainly through the cross-recombination of the electrons and OH radicals generated by the ionization of water (that occurs with a rate constant of 3x10$^{10}$ M$^{-1}$ s$^{-1}$). The path-average concentration of the electrons is 1.6 mM, which gives a time constant of 21 ns; the observed time constant is higher, ca. 5 ns, due to the extremely non-homogeneous excitation profile for the "3+1" photoprocess.

**Fig. 3.**

(a) Normalized FDSS kinetics obtained for three- 400 nm photon excitation of N$_2$-saturated liquid water flowing in a high-speed jet *(solid line)*. Empty circles indicate the pump probe kinetics obtained under the same excitation conditions. The maximum *ΔOD* is ca. 0.01; the compression factor $s=-2160$ and $\tau_{GVD}=0.9$ ps (1 ps = 6.5 cm$^{-1}$). An arrow indicates the "spike" in the oscillation pattern near the kinetics origin that is analogous to the same "spike" in the FDSS kinetics obtained for biphotonic detachment of the electron from aqueous iodide (ref. 3). (b) Power dependence for normalized FDSS kinetics obtained under the same "3+1" excitation conditions as the PPS kinetics shown in Fig. 2. The pump power and the maximum $\Delta OD$ were (i) 27 µJ and 0.017, (ii) 49 µJ and 0.2, and (iii) 94 µJ and 0.394, respectively. Note the drastic reduction in the oscillation amplitude at the higher optical density. The FDSS kinetics were obtained for $\tau_{GVD}=1.2$ ps (1 ps = 3.6 cm$^{-1}$) with a 33 fs FWHM (seed) probe pulse. Traces (i), (ii), and (iii) are the averages of 30, 45, and 100 thousands shots, respectively.

**Fig. 4.**

Pump-probe kinetics observed upon the 400 nm excitation of 1.3 µm film of amorphous hydrogenated silicon (*a*-Si:H) on a suprasil substrate (801.3 nm probe). The pump power was (i) 9 and (ii) 43 µJ. After the initial rapid decay on the picosecond time scale due to carrier relaxation; a slower decay on sub-nanosecond time scale is due to the recombination of trapped charges in the bulk. The 33 fs FWHM probe pulse is centered at the transmission extremum (see Fig. 5) and has a band pass similar to the fringe spacing. Note the flatness of the decay kinetics for $t>10$ ps at the lower excitation power.

**Fig. 5.**

(a) FDSS "kinetics" $\Delta OD$ for the *a*-Si:H system (see caption to Fig. 4) plotted as a function of $\Delta\omega$. These kinetics were obtained with a 400 nm pulse of 19 µJ and (i) $T=9$ ps and (ii) $T=0$ ps, respectively. The compression factor $s$ for a 33 fs FWHM probe pulse is -630, $\tau_{GVD}=0.53$ ps (1 ps = 19 cm$^{-1}$). The spectral response function is given by the $T=300$ ps trace (iii). (b) The kinetics from Fig. 5(a) normalized by the spectral response function $T'(\omega)$ (plotted to the right and to the top).

**Fig. 6.**

(a) Same as Fig. 5(b), for a different spot on the same a-Si:H sample and greater GVD. FDSS traces (i) and (ii) were obtained for (i) $T=50$ ps and (ii) $T=0$ ps, respectively. Trace (iii) (the spectral response function plotted to the right and to the top) was obtained at $T=300$ ps. The maximum $\Delta OD$ was ca. $6\times10^{-2}$. The kinetics were obtained for a compression factor of $s = -3780$ and $\tau_{GVD}=1.23$ ps (1 ps = 3.6 cm$^{-1}$). (b) The kinetic data of Figs. 5(b) and 6(a) replotted vs. the "absolute" reduced time, $(T_e - T)/\tau_{GVD}$. FDSS kinetics at $T=0$ ps (lines with symbols) and T= 9 ps (for trace (i)) and 50 ps (for trace (ii)) (dotted lines) are shown. For traces (i) s=-630 and $\tau_{GVD}=0.52$ ps, for traces (ii) s=-3780 and $\tau_{GVD}=1.23$ ps, respectively. The oscillation pattern does not change with the GVD or the delay time of the pump. Compare this figure with Figs. 7 and 8.

**Fig. 7.**

FDSS kinetics $S(\omega)$ given by eq. (3) for a thin-film sample with a complex index of refraction $n_\omega = 3.44 + 0.0011i$ [14] and $d=1.3$ µm, for three pump delays: $T=0$ ps (i), 20 ps (ii), and 35 ps (iii). The simulation parameters were $\phi_\varepsilon=90°, \tau_p=20$ fs, s=2048, $1/e$ width of the pump pulse of 100 fs, and the lifetime of photogenerated species of 40 ps; we let $|\Delta\varepsilon_\omega| = 1$. Traces (a) and (b) (to the right) are the spectral response functions $T'(\omega)$ for pure photoabsorption, $\phi_\varepsilon=90°$, and photorefraction, $\phi_\varepsilon=0°$, respectively. For photoabsorption ($\phi_\varepsilon=90°$), the function $T'(\omega)$ has the same extrema as the sample transmission $T_\omega \propto |t_\omega|^2$. For photorefraction ($\phi_\varepsilon=0°$), $T'(\omega)$ has zeroes at the transmission extrema and maxima and minima at the inflection points of $T_\omega$.

**Fig. 8.**

(a) Same as Fig. 7, but simulated kinetics $S(\omega)$ were normalized by the spectral response function $T'(\omega)$. For long group delay times $T_e$ these normalized kinetics asymptotically approach the TA kinetics $\Gamma(t)$. (b) A comparison between the normalized oscillation patterns obtained for different delay times $T$ of the pump (the latter are given in the legend in the plot): $T=0$ (solid line), $T=20$ ps (filled circles), and $T=35$ ps (empty squares). The dashed line indicates zero signal.

Figure 1; Shkrob et al.

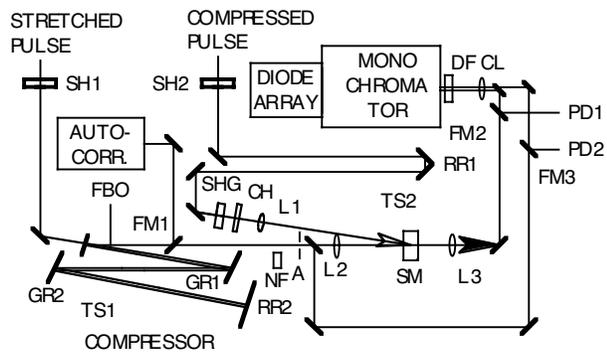

Figure 2; Shkrob et al.

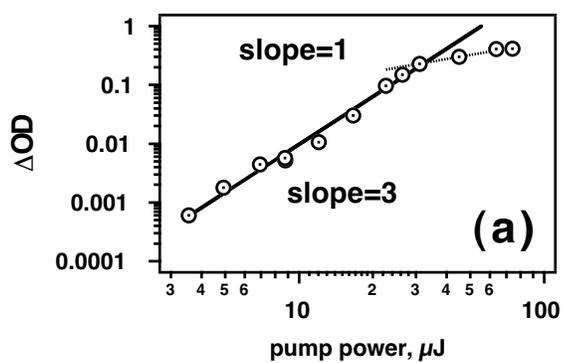

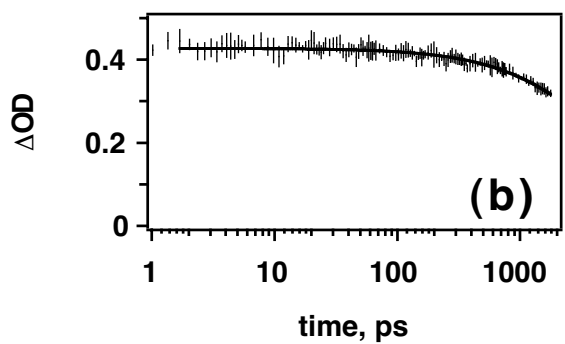

Figure 3; Shkrob et al.

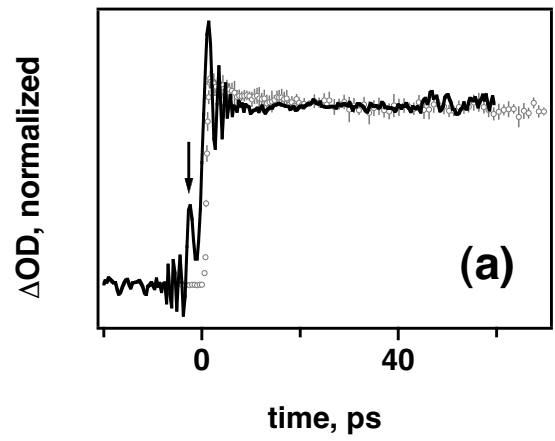

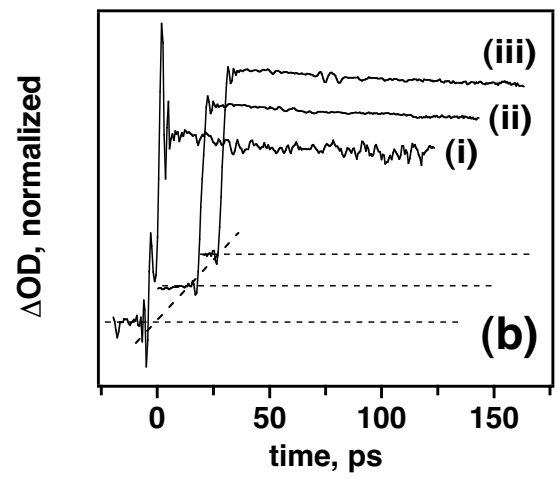

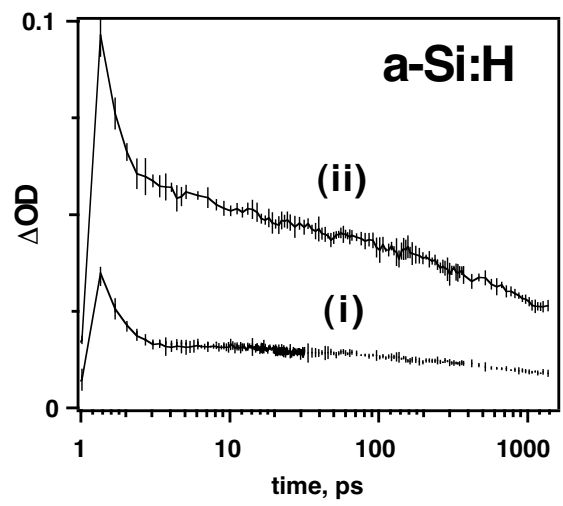

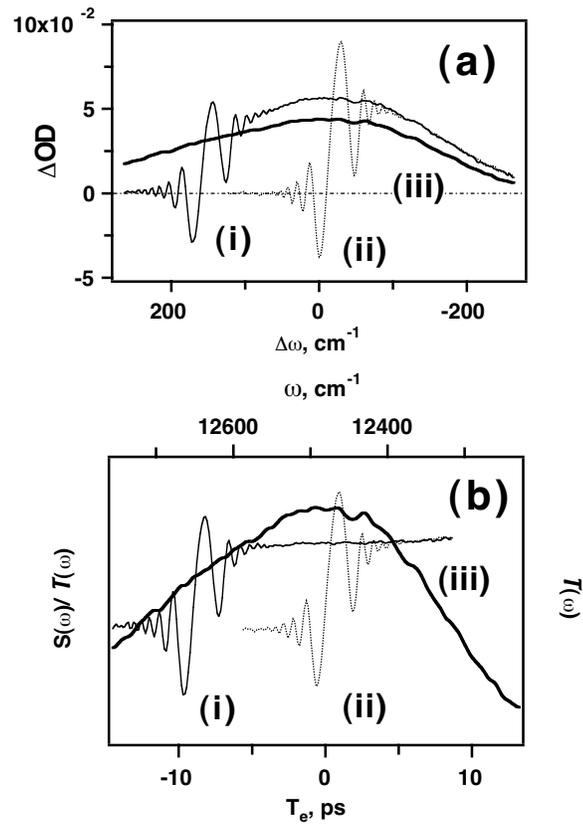

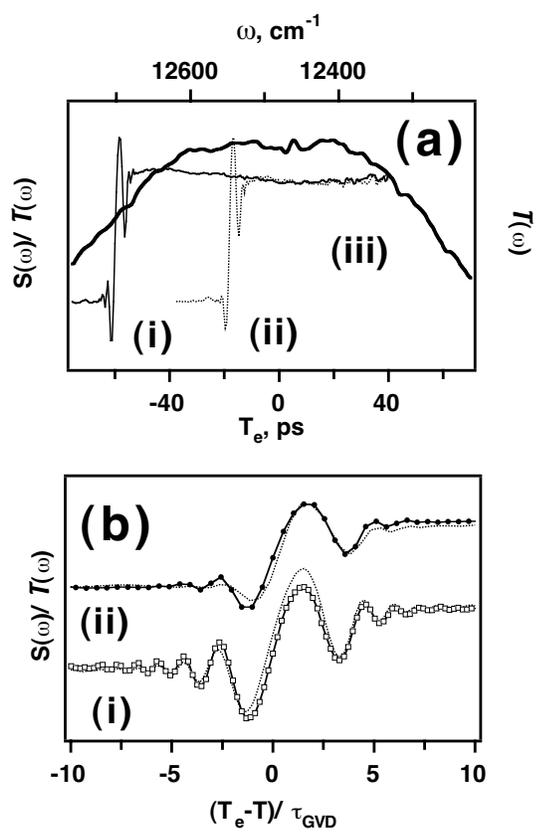

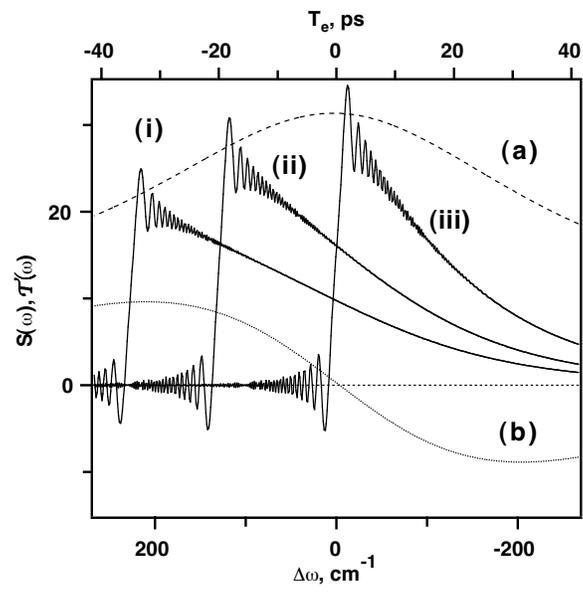

Figure 8; Shkrob et al.

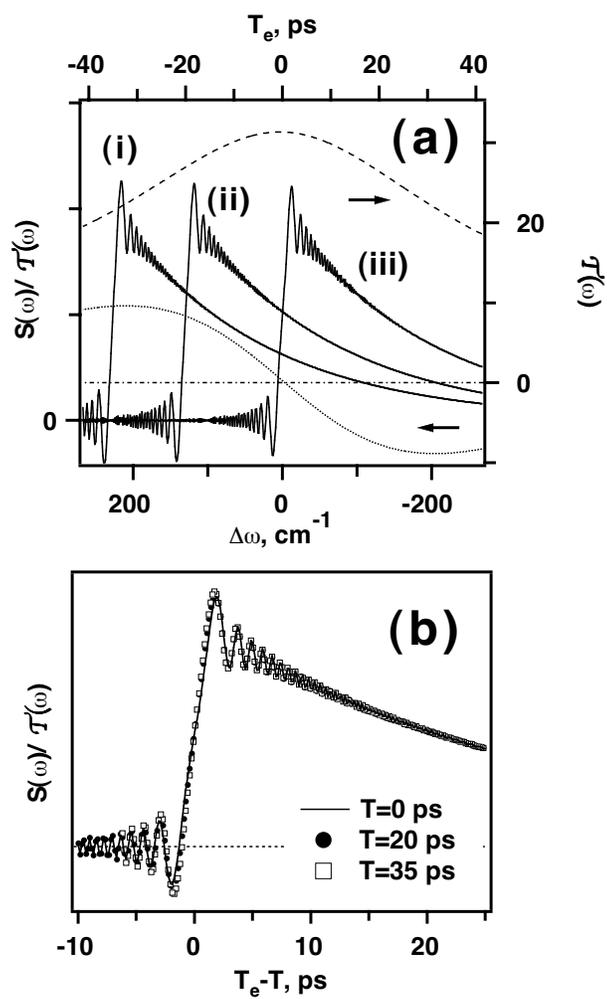